# Theoretical study of the stable states of small carbon clusters $C_n$ (n = 2-10)

D. P. Kosimov[1], A. A. Dzhurakhalov[1,2], and F. M. Peeters[1]

[1]*Department of Physics, University of Antwerp, Groenenborgerlaan 171, B-2020 Antwerpen, Belgium*

[2]*Theoretical Dept., Arifov Institute of Electronics, F. Khodjaev Street 33, 100125 Tashkent, Uzbekistan*

**Abstract.** Both even- and odd-numbered neutral carbon clusters $C_n$ (n = 2-10) are systematically studied using the energy minimization method and the modified Brenner potential for the carbon-carbon interactions. Many stable configurations were found and several new isomers are predicted. For the lowest energy stable configurations we obtained their binding energies and bond lengths. We found that for n ≤ 5 the linear isomer is the most stable one while for n > 5 the monocyclic isomer becomes the most stable. The latter was found to be regular for all studied clusters. The dependence of the binding energy for linear and cyclic clusters versus the cluster size n (n = 2-10) is found to be in good agreement with several previous calculations, in particular with *ab initio* calculations as well as with experimental data for n = 2-5.

## I. INTRODUCTION

The properties and structures of carbon clusters have attracted a lot of attention from both physicists and chemists over the last fifty years. Several of these theoretical and experimental studies have focused on the characterization of these clusters (see e.g. the review papers[1,2]). The unique properties and the reduced dimensionality of small carbon clusters are promising for various potential applications. The recent discovery of single graphene layers[3] has revived the interest in low dimensional carbon clusters and carbon-based nanoribbons.

Theoretical studies on carbon clusters can be divided into two main categories: calculations using *ab initio* techniques[4-8] and calculations based on empirical interatomic potentials.[9-12] Accurate *ab initio* calculations performed by Raghavachari and Binkley[5] obtained the structures and energies of small carbon clusters ($C_n$, n = 2-10) and predicted an odd-even alternation in the nature of the cluster geometries with the odd-numbered clusters having a linear structure and the even-numbered clusters preferring an irregular cyclic structure. Several calculations have confirmed such an odd-even behavior of these clusters.[12-15] Jones[14] performing density functional theory (DFT) calculations for even neutral carbon clusters ($C_n$, n = 4-32) found several stable isomers including chains, rings, cages, and graphitic ("plate" and "bowl") structures, and showed the difference in binding energies of the same structural type as obtained by local spin density (LD) and gradient-corrected (Becke-Perdew: BP) approximations. In these calculations the cyclic structures (rings) were irregular only for n ≤ 8.

Along with quantum chemical calculations of carbon clusters, extensive semi-empirical tight-binding (TB) methods were used to obtain equilibrium geometries for carbon clusters of arbitrary size. In some cases the cohesive energies[16] and the stable final equilibrium geometries[17] of small carbon clusters obtained by TB are in good agreement with *ab initio* calculations. However, although the odd-even alternation was also observed in tight-binding molecular dynamics (TBMD) calculations,[9] the cyclic structures were rather regular. A genetic algorithm using the Brenner bond-order potential[18] was applied to minimize the energy of $C_n$ (n=2-60[10], n=2-30[11]) clusters. The ground-state structure was found to evolve from linear chains (n=2-4[10], n=2-5[11]) to monocyclic rings (n=5-17[10], n=6-12[11]) to polycyclic rings and to fullerenes. Recently, the configurations and energetics of small carbon clusters as obtained from different types of TB and local density approximations were compared, and a difference in the configuration of even numbered clusters was predicted by short and long range TB models.[19]

In spite of the extensive studies of carbon clusters a systematic study of the possible stable configurations of small clusters within a single approach is still lacking. Moreover, the conclusions of the different methods do not always agree with each other. Up to now all previous studies were mainly limited to a calculation of the ground state or at most a few stable isomers. Here we will go beyond such approaches and obtain many more stable isomers of clusters for n up to 10 within the same theoretical method. The motivation is that in nature finite size systems do often not reach their ground state when they are formed and may get stuck in a metastable state. We will compare systematically the configurations, bond lengths and binding energies of clusters predicted in our calculation with available previous theoretical results and with experimental data.

This paper is organized as follows. In Sec. II, the basic idea of the local minimum energy method and a description of the used potential in our calculation are presented. The geometries and energetics of many possible isomers of small carbon clusters $C_n$ (n = 2-10) predicted in this study are compared with available data of others in Sec. III. In Sec. IV, the results of the present study are summarized.

## II. SIMULATION METHOD

Energy minimization methods are common techniques to compute the equilibrium configuration of molecules and clusters. The basic idea is that a stable state of a molecular system should correspond to a local minimum of their potential energy.[20] The lowest energy state corresponds to the ground state while the other local minima are metastable states.

Since *ab initio* quantum chemical methods are extremely CPU-time consuming, we decided to use the classical interatomic potential approach. A very accurate approach is based on the empirical potential which was proposed by Brenner[18] and is based on the Tersoff bond-order expression that was fitted on diamond, graphite crystal and small hydrocarbons. This method is orders of magnitude faster than *ab initio* schemes. It was shown that this potential can describe accurately not only the different bulk condensed phases (e.g., diamond and graphite), but also turned out to be reliable for small systems as carbon clusters which was demonstrated in many works.[10,11,21-24] In particular, Hobday and Smith[10] optimized the structure of $C_n$ (n = 3–60) using the Brenner potential by implementing a genetic algorithm. In Ref. [23] the structures with the lowest energies of the carbon clusters $C_n$ with n from 2 to 71 were successfully obtained using a global optimization algorithm with the Brenner potential for the generation of carbon cluster structures.

In the present work we use for the carbon-carbon interaction potential the recently modified version of the Brenner potential.[25] Among various bond order potentials, this so called Brenner second-generation reactive empirical bond order (REBO) potential is parameterized specifically for carbon and hydrocarbon systems. This potential is based on the empirical bond order formalism and allows for covalent bond binding and breaking with associated changes in the atomic hybridization. Consequently, such a classical potential allows to model complex chemistry in large many-atom systems. The Brenner bond order potential can be written in the following general form for the binding energy

$$E_b = \sum_i \sum_{j(>i)} [V^R(r_{ij}) - b_{ij} V^A(r_{ij})].$$

The first term is repulsive and the second one is attractive. $r_{ij}$ is the distance between pairs of nearest-neighbor atoms *i* and *j*. Although this expression is a simple sum over bond energies, it is not a pair potential since the $b_{ij}$, which is called the bond order factor, is in essence a many-body term. For example, in the Brenner potential the $b_{ij}$ depends on the bond and torsional angles, the bond lengths and the atomic coordination in the vicinity of the bond. The many-body nature of $b_{ij}$ makes the bond energy to depend on the local environment of the bond. This feature allows the Brenner potential to predict correct geometries and energies for many different carbon structures and the correct hybridization. The values for all the parameters used in our calculation for the Brenner potential can be found in Ref. [25] and are therefore not listed here.

In our study the initial coordinates of the cluster atoms are taken to be random in a given area. This area is planar for the generation of planar structures and three dimensional for a generation of non-planar ones. The energy minimization of each cluster is performed on typically 500 different random configurations. Starting from each initial configuration the final configuration is found by minimizing the energy of the system. For the initial planar geometries we also allowed the atoms to move out of the plane. If the energy and the configuration of the obtained minimum are different from the previous ones, this configuration is considered as new. When the final bond lengths between given atom (or an ensemble of few atoms) and all other atoms are larger than 2 Å, this configuration is disregarded.

## III. RESULTS AND DISCUSSION

In this paper we present the most important lowest energy stable configurations with their bond lengths for neutral carbon clusters $C_n$ with $n \leq 10$. For all studied clusters the different structures were found and the number of possible stable configurations increases drastically with increasing n. In Figs. 1-6 the isomers of a given cluster are ordered and labeled according to their energies: e.g. 5(1) corresponds to the lowest energy form of $C_5$, 5(2) is the second lowest energy configuration, etc. We now turn to a detailed description of the found structures.

### A. $C_2$ and $C_3$

In the case of $C_2$ only one trivial linear configuration is possible (see Fig. 1). However, the bond length depends slightly on the calculation method. The *ab initio* calculations of Refs. [5, 8] and experiment[26] gave a bond length of 1.245 Å for the C-C bond while 1.31 - 1.34 Å was obtained in Ref. [27]. Other authors reported for this bond length the values 1.413 Å[9] and 1.315 Å.[11] In our calculation the bond length for $C_2$ was found to be 1.325 Å which corresponds to a typical C-C double bond. In this geometry the binding energy of $C_2$ is -6.2098 eV which corresponds to 3.10 eV per atom. Menon et al.[17] obtained 2.82 eV per atom using the generalized TBMD. Note that our result is identical to the experimental found value for the binding energy per atom as given in Ref. [26] which is 3.10 eV.

For $C_3$ the lowest energy linear structure (-12.9018 eV or 4.30 eV per atom) calculated in our study has a bond length $r_{12} = r_{23} = 1.310$ Å which is in good agreement with data from others.[11,27] Menon et al.[17] reported 4.85 eV per atom for this structure. The experimental value for the binding energy per atom is 4.63 eV[28] and 4.57 eV.[29] The next stable structure for $C_3$ is a regular triangle (cyclic) configuration (-9.2847 eV) with C-C-C bond angle of $60^0$ (see Fig. 1) and bond length of 1.617 Å (single bond) which is significantly different from 1.363 Å as found in Ref. [8]. In our case all other irregular triangle structures of $C_3$ have higher binding energy than the regular one although Menon et al.[17] found the most stable triangle configuration with lower symmetry which has a bond angle $98^0$ and a bond length 1.37 Å.

## B. $C_4$

For clusters with n ≤ 5 our approach with the Brenner potential gives the linear structure as the ground state configuration which is in good agreement with results of Refs. [11, 14]. In the triplet linear form of $C_4$ we found the outer bond length to be $r_{12} = r_{34} = 1.310$ Å and the inner bond length is $r_{23} = 1.331$ Å. The binding energy is -19.0187 eV (or 4.75 eV/atom) which is in very good agreement with experiments (-19.0 eV),[28] (19.4 eV)[29] and other calculations (4.80 eV/atom).[17] Although the bond lengths are close to those found by others, the size of the inner and outer bond lengths were sometimes found to be interchanged. Some authors[5,14,27,30,31] reported that the inner bond length is shorter than the outer ones, while others[9,11,17] found the opposite in agreement with our results.

The other possible geometries for the $C_4$ cluster are shown in Fig. 1. The next stable structure 4(2) with a binding energy of -16.9292 eV and bond length of 1.364 Å is three-fold symmetric. We did not find this structure in the literature. A modified version of it was found, i.e. a branched structure 4(3) with binding energy -15.4275 eV was obtained earlier.[8,32] However, the triangular part of the structure 4(3) has a single bond (1.603 – 1.636 Å) and the outer atom has a double bond (1.319 Å). Whiteside et al.[32] reported that this structure is less stable than the following discussed structure 4(4). The next stable structure (-15.0022 eV) of $C_4$ is a cyclic (rhombic) isomer with a bond length of 1.567 Å and acute angle of $78.5^0$. This bond length is close to the one (1.530 Å) with the acute angle of $63.2^0$ found in Ref. [31]. In Refs. [5] and [17] these values are 1.425 Å, $61.5^0$ and 1.40 Å, $69.6^0$, respectively. Some authors[5,17,32] have confirmed that the rhombic form of $C_4$ is slightly more stable than its linear form, others found that these two isomers are essentially isoenergetic[33] and that the linear structure is the most stable.[9,12,14]

The square form with the binding energy -14.6776 eV and bond length of 1.546 Å is the fifth isomer. The non-planar (tetrahedral) configuration 4(6) is the next form of $C_4$ with binding energy and bond length of -13.0821 eV and 1.721 Å, respectively.

## C. $C_5$

The possible stable $C_5$ clusters obtained in our studies are also depicted in Fig. 1. For the linear form of the $C_5$ cluster, as mentioned above, the outer bond length (1.310 Å) is shorter than the inner bond lengths (1.331 Å) which are in good agreement with the results 1.323 Å and 1.332 Å, respectively of Ref. [11]. We found that the values of the bond lengths for such linear structures are the same for all clusters with $4 \leq n \leq 10$. The linear structure has a binding energy of -25.1357 eV (5.03 eV/atom) close to the one (4.91 eV/atom) found in Ref. [11] and slightly higher than 5.38 eV/atom obtained in Ref. [17]. The experimental value is 5.30 eV/atom.[28]

Our calculation predicts the existence of a regular cyclic (pentagonal) configuration 5(2) of $C_5$ with the bond length of 1.437 Å, interior angles $108^0$ and binding energy -23.7099 eV (4.74 eV/atom). Note all cyclic configurations studied by us are regular which is in agreement with Ref. [9] for n-even (n = 6, 8, 10) clusters and with Ref. [11] for $6 \leq n \leq 12$ clusters. Irregular cyclic stable forms with lesser symmetry were also reported for n-even (n = 4-10),[5] $n \leq 10$ (except n = 7, 9 which were unstable),[17] n = 4, 6, 8[14] and n-odd (n = 5-15)[7] clusters.

The structure 5(3) with binding energy -22.9388 eV is the evolution of 4(2) formed by adding one more atom. Its further evolution is also observed for n > 5 forming its longer linear part. The outermost atom in its linear part has a bond length 1.310 Å which is similar as in the case of linear isomers. This atom results in an increasing bond length (1.399 Å) for the next two atoms as compared to the one of three-fold symmetric configuration.

The next two structures are a modification of the branched structure 4(3): the configuration 5(4) has a binding energy of -21.7913 eV and the isomer 5(5) lies 0.4376 eV higher in energy than 5(4). Notice that the bond length of the triangle part is almost the same as in the cyclic form of $C_3$. The outer atoms have a double bond. The structures 5(6) and 5(7) have a similar tetragonal basis with different attachments. This difference results in different binding energies and bond lengths in these structures: -20.9252 eV with bond length 1.329 Å for the attached atom and -20.6243 eV with 1.343 Å for the attached atom, respectively. Besides, their tetragonal part is less symmetric than the tetragon 4(4): the bond lengths are 1.599 Å, 1.561 Å in its attached and free parts, respectively for 5(6), and 1.624 Å, 1.515 Å, respectively for 5(7).

The next three [except 5(10)] configurations are non-planar structures. First, the regular diamond-like structure 5(8) is predicted with total energy of -19.1919 eV and bond length of 1.432 Å, although its irregular form is also found with a slightly (0.14 eV) lower energy. In the later case four carbon atoms form tetrahedron with bond length 1.703 Å in the base plane and 1.730 Å in the direction of the apex, and fifth atom with bond length 1.329 Å is attached to this apex. The square pyramidal form 5(9) of $C_5$ has energy -17.9409 eV while its triangle dipyramidal form 5(11) has a slightly higher energy: -17.8634 eV. In the square pyramidal structure the bond length is 1.695 Å in the base plane and 1.738 Å in the direction of the apex. The triangular dipyramid has a bond length 1.821 Å at the three edges of a common base plane and 1.707 Å at the other six edges. The central atom of the structure 5(10) with binding energy -17.8833 eV has the bond length of 1.701 Å and the bond length of the two neighbor pairs is 1.514 Å. This structure has an isomer with binding energy -16.6012 eV and bond length 1.509 Å with the only central atom. Similar structures to 5(4) and 5(9) were also obtained recently for the negative charged state of $C_5$,[34] and to 5(6) for the negative[34] and positive[35] charged states of $C_5$.

## D. $C_6$

Fig. 2 shows the most stable geometries with their corresponding binding energy for $C_6$. Note that now the linear form has a energy 0.4691 eV higher than the cyclic (or ring-like) one which becomes the ground state for n ≥ 6. This is the famous benzene ring. The binding energy of this cyclic structure 6(1) is -31.7217 eV (5.29 eV/atom) with a bond length of 1.391 Å.

The structures from 6(3) to 6(11) [except 6(4)] are various evolutions of the above considered and discussed geometries of $C_5$. The fourth most stable structure 6(4) of $C_6$ is the pentagon with one extra carbon atom attached to it which has a binding energy of -29.0040 eV. Its further modification formed by attaching external atoms, which is called a "tadpole" structure, will also be considered for n > 6. The bond length for the external atom is 1.348 Å and in the pentagon $r_{15} = r_{54} = 1.485$ Å, $r_{12} = r_{34} = 1.425$ Å, and $r_{23} = 1.421$ Å.

The structures 6(12)-6(18) except 6(16) are non-planar. These results show that the most non-planar structures have an energy substantially larger than the ground state. The structure 6(12) is a new form of $C_6$ in which two atoms are situated out of the plane defined by the other four atoms. These two atoms form triangles with bond length of ~1.70 Å. Other four atoms have bond length 1.422-1.464 Å. Modified versions of this structure were also found for n > 6.

The pentagonal pyramid 6(14) has bond length 1.615 Å in the basal plane and 1.788 Å in the direction of the apex. The square dipyramidal form 6(17) with a bond length 1.728 Å at all its edges has 1.9288 eV higher energy than 6(14). The triangular prism form 6(15) of $C_6$ has bond length 1.705 Å in the basal plane and 1.666 Å between the two bases. The fish-like structure 6(16) is the less stable form of $C_6$ as compared to the other planar configurations. It has bond lengths $r_{12} = r_{13} = 1.533$ Å, $r_{24} = r_{34} = 1.707$ Å and $r_{45} = r_{46} = 1.478$ Å.

### E. $C_7$, $C_8$, $C_9$, $C_{10}$

The found stable geometries for $C_7$ - $C_{10}$ clusters are depicted in Figs. 3-6. As n ≥ 6 (mentioned above), for these clusters the ground state is the ring-like structure. The bond length in this structure decreases slightly from 1.374 Å to 1.352 Å with increasing n. The next stable configuration is the linear structure for $C_7$ and $C_8$, and then it becomes the third (for $C_9$) and the fifth (for $C_{10}$) most stable configuration. For this isomer, no changes are observed in bond length (i.e. the outermost one is 1.310 Å and all inner bond lengths are 1.331 Å), though the total binding energy of these clusters as well as its energy per atom becomes lower with increasing n.

Notice that for each n ≥ 4 cluster, new n-1 cyclic configurations with one external attached atom are found to be stable. Therefore the different cyclic configurations with one or more attached atoms, i.e. "tadpole" structures, are observed from 3 to n-1 for each n ≥ 4. For $C_7$ this new cyclic configuration is 7(3) (see Fig. 3). The structures 7(4)-7(14) are evolutions of corresponding structures in $C_6$ that are formed by adding one more atom. Comparison of 7(5), 7(7) and 7(9) shows that the binding energy is lower, if external atoms are attached to the same atom or neighbor atoms of the cyclic core. The structures 7(15) and 7(16) are further evolutions of the diamond-like structure 6(13). The next two structures are the hexagonal pyramid 6(17) with bond length 1.561 Å in the basal plane, 1.823 Å at the apex, and the pentagonal dipyramid 6(18) with bond length 1.724 Å in the basal plane, 1.750 Å at the apex. The structure 7(19) is planar with bond length 1.720 Å.

Most of the configurations of $C_8$ were found to be similar to the above discussed ones (see Fig. 4). However, the number of various combinations of such structures and the number of non-planar structures increases in this case. Besides the seeds for the planar graphene structure can be found already for $C_8$ as the double ring structure consisting of two pentagons 8(8). This becomes a penta-hexagon double ring in the case of $C_9$ and actual double hexagons in the case of $C_{10}$. Some buckled cyclic structures are also possible that can be observed for n ≥ 6 [e.g. 8(14), 8(15), 8(24) and 6(12)]. Further increase of the linear part in the diamond-like structure 7(16) was not found in case of the $C_8$ - $C_{10}$ clusters, however, one can see variations to it with increasing n: the structures 8(23), 8(25) (Fig. 4), 9(26), 9(27) (see Fig. 5), 10(15), 10(27) and 10(31) (see Fig. 6). Our calculation predicts the cubic 8(27) and hexagonal dipyramidal 8(28) geometries of $C_8$ as being almost the least stable structures. However, Jones[14] found the cubic isomer with bond length 1.468 Å as the third stable after the ring and linear configurations. In our case the cube has a bond length of 1.669 Å. Notice that the pyramidal and dipyramidal form for $C_9$ and $C_{10}$ were not found as stable configurations in our energy minimization.

As shown in Figs. 5 and 6, some modifications of cyclic configurations lie between the most stable ring and the linear structures of $C_9$ and $C_{10}$. In the case of $C_9$ there is only one cyclic configuration with one attached atom between these structures, and there are three such cyclic structures for $C_{10}$. The hexagonal double ring 10(2) becomes the second most stable structure. Jones[14] reported it as the third stable after the ring and linear geometries. The pentagonal prism 10(30) which is rather far from the ground state in our calculation was found as the fourth stable isomer in Ref. [14].

Various combinations of structures similar to 8(29), 9(25) and 10(32) where an external atom is attached to two atoms of the cyclic isomer were obtained for all n > 4 clusters, which are not shown. Another case is that when one cyclic (or other) structure is attached to the rest with one atom [see, e.g. 10(20)]. Such structures are not shown in the figures.

Notice that similar structures as the non-planar 5(9), 8(27) and 10(30) ones were also obtained recently for the charged state of $C_5$.[34,35] Similar structures to 6(9), 7(6) and 9(4) were obtained recently for the negative charged state[34] but with a lesser symmetry. The positive charged state of $C_8$ has the cyclic structure as its ground state,[35] while for the negative charged state the linear structure becomes the ground state.[34]

**F. Linear and cyclic structures: comparison**

We now turn to the comparison of the two most stable, i.e. linear and cyclic, configurations. The binding energy per atom and the bond length of the linear and cyclic $C_n$ as obtained in the present approach are depicted in Fig. 7 as a function of cluster size n. The binding energy for both structures increases (in absolute value) steeply at first and then slowly with increasing n [see Fig. 7(a)]. Our results are compared with previous Hartree-Fock (HF)[5] calculations. Notice that while the linear form is favorable for very small clusters, the cyclic form becomes more stable when n ≥ 6. Our calculated binding energies are in good agreement with the HF results, but our results have a slightly lower energy. The bond length for the linear configuration almost does not change as function of n, while the bond length of the cyclic one decreases with increasing n approaching the value for linear structures in cases of large clusters [see Fig. 7(b)].

In Fig. 8 our results for the binding energy of the two most stable forms are compared with other calculations and with available experimental data. The experimental data shown in Fig. 8(a) for the linear form with n = 2,[26] and n = 3, 4, 5[28] show that our theoretical results are very close. The binding energy for the linear structures calculated by Tománek and Schluter[12] using a combination of an adaptive simulated method and a simple TB-type Hamiltonian for the total energies has n-even-odd oscillations and higher absolute values for the binding energy than found experimentally and found from the present work. Their data as shown in Fig. 8(b) for the cyclic isomer show that this structure is less stable than the linear one having almost the same energy for n = 6 and n = 10. Using TBMD simulations the n-even-odd oscillation has also been found by Xu et al.[13] for both the linear and cyclic forms. Besides they confirmed that in the range $5 \leq n \leq 11$ the odd-numbered clusters prefer a linear structure, while even-numbered clusters prefer a cyclic structure as reported also in Ref. [5]. However, a local spin density (LD) calculation made in Ref. [14] shows that a cyclic structure prevails in energy above the linear one for $n \geq 6$ which agrees with the present results. This conclusion is supported by Refs. [10, 11] where a genetic algorithm for the optimization with a Brenner potential was used.

## IV. CONCLUSION

The most stable structures of neutral carbon clusters $C_n$ (n = 2-10) were found using an energy minimization method with a modified Brenner potential. Many new isomers were found. We ordered the isomers according to their energies. In the present study we found that the linear structure is the most stable one for clusters with $n \leq 5$ and the monocyclic form is the most stable for $n \geq 6$ which are in good agreement with several previous theoretical calculations.[10,11,14] The monocyclic isomer in our case is regular for all clusters studied. *Ab initio* calculations found also an irregular monocyclic form for these clusters. It was shown that most non-planar structures have a higher energy, i.e. for small clusters the favorable geometries are planar structures. The found seeds for the planar graphene structure in the case of $C_8$ - $C_{10}$ can be useful for the understanding and the study of submonolayer graphene.

The binding energy for linear and cyclic clusters was calculated versus the cluster size which was in good agreement with results of earlier calculations and with available experimental data for n = 2-5. We did not find any n-even-odd oscillations for linear and cyclic isomers of the studied clusters.

**Acknowledgments**

This work was supported by the Belgian Science Policy (IAP) and the Flemish Science Foundation (FWO-Vl).

**Figure captions**

FIG. 1. (Color online) The stable configurations for the isomers of $C_2$, $C_3$, $C_4$ and $C_5$.

FIG. 2. (Color online) The most stable geometries for the isomers of $C_6$. The total binding energies in eV are shown above the corresponding structures.

FIG. 3. (Color online) The same as Fig. 2, but now for $C_7$.

FIG. 4. (Color online) The same as Fig. 2, but now for $C_8$.

FIG. 5. (Color online) The same as Fig. 2, but now for $C_9$.

FIG. 6. (Color online) The same as Fig. 2, but now for $C_{10}$.

FIG. 7. (Color online) Binding energy $E_{coh}$ of carbon clusters with linear (L) and cyclic (C) structures calculated in present study and by HF[5] as a function of cluster size n (a) and their bond length (b).

FIG. 8. (Color online) Comparison of the binding energy $E_{coh}$ of carbon clusters with linear (a) and cyclic (b) structures as a function of cluster size n obtained using TB,[12] BP, LD,[14] TBMD[13] calculations and measured in experiments for n = 2,[26] and n = 3, 4, 5.[28]

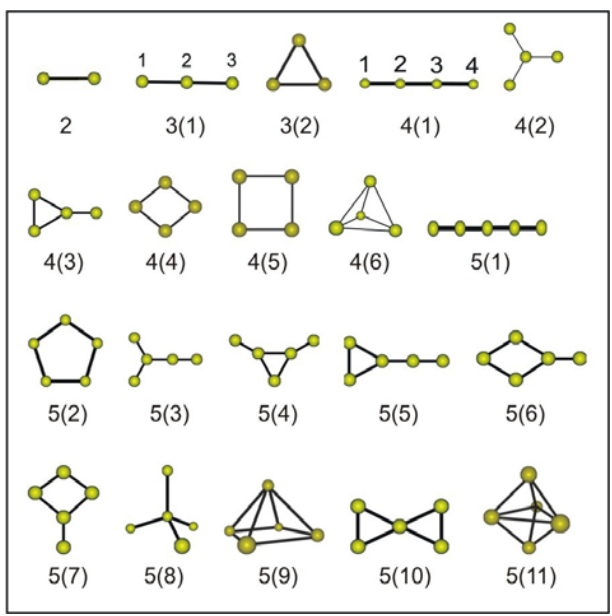

**FIG. 1**

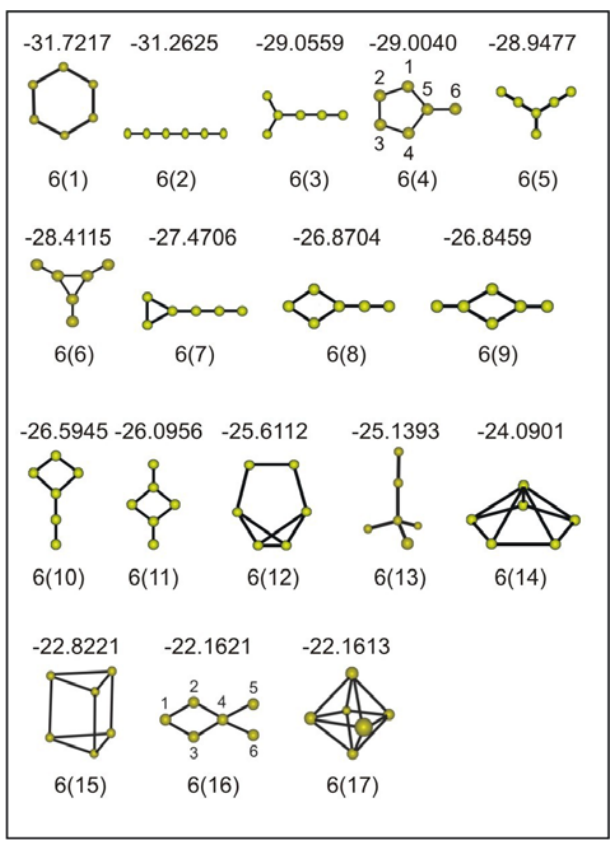

**FIG. 2**

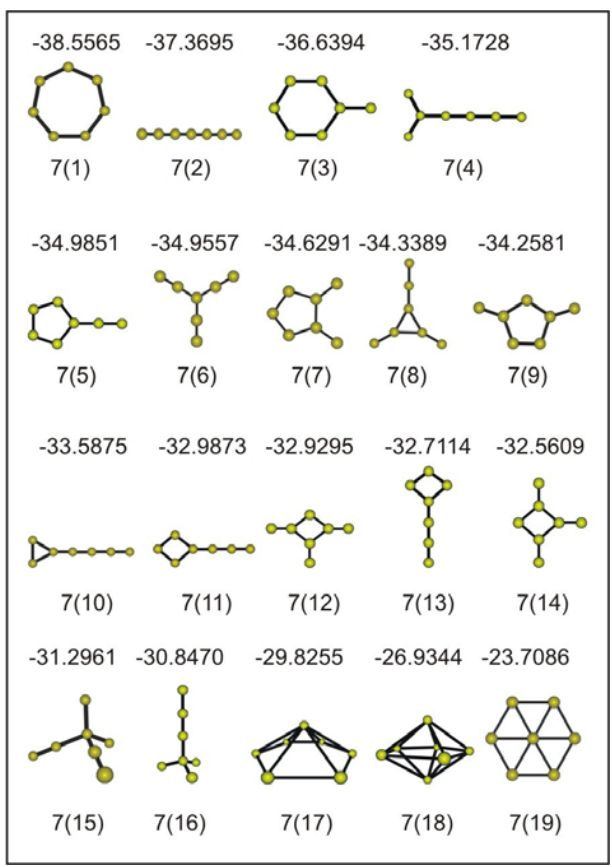

**FIG. 3**

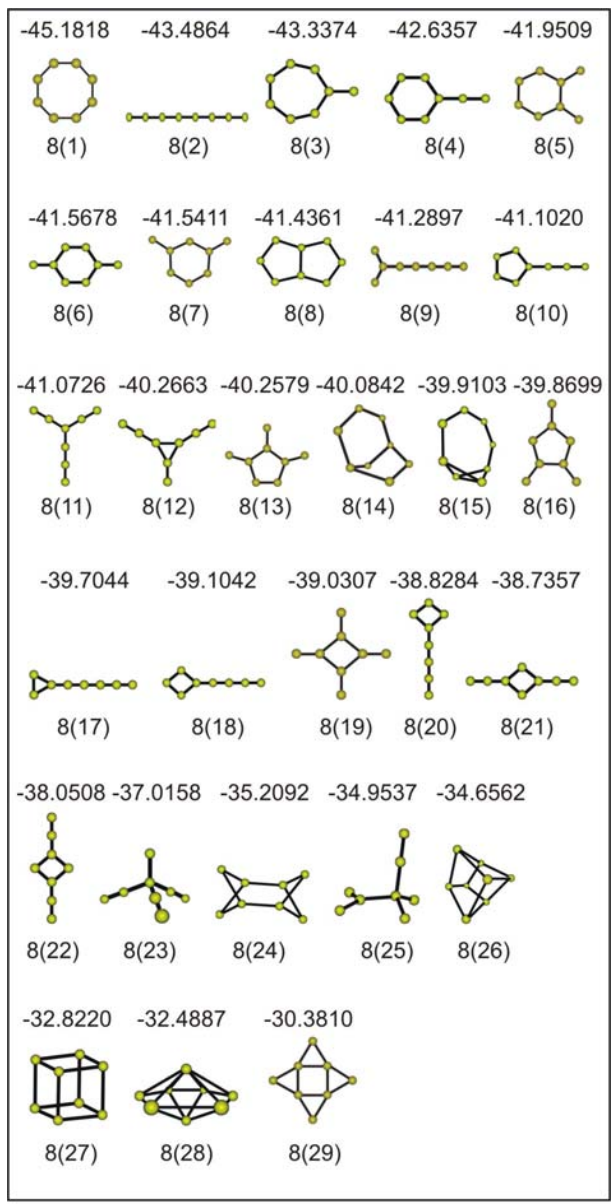

**FIG. 4**

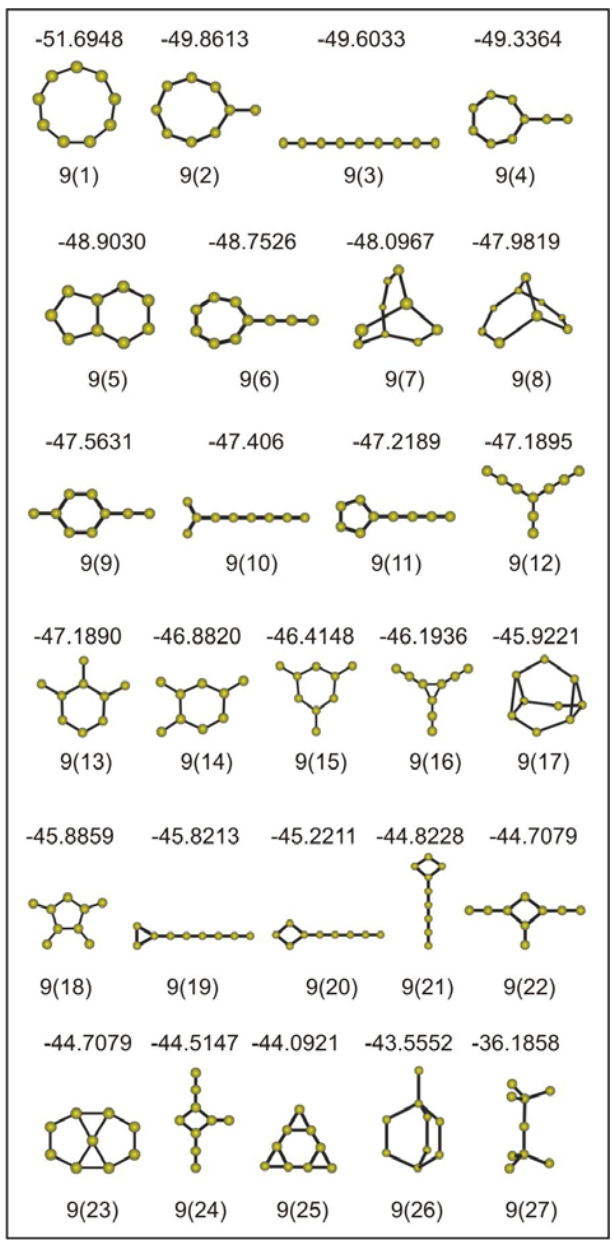

**FIG. 5**

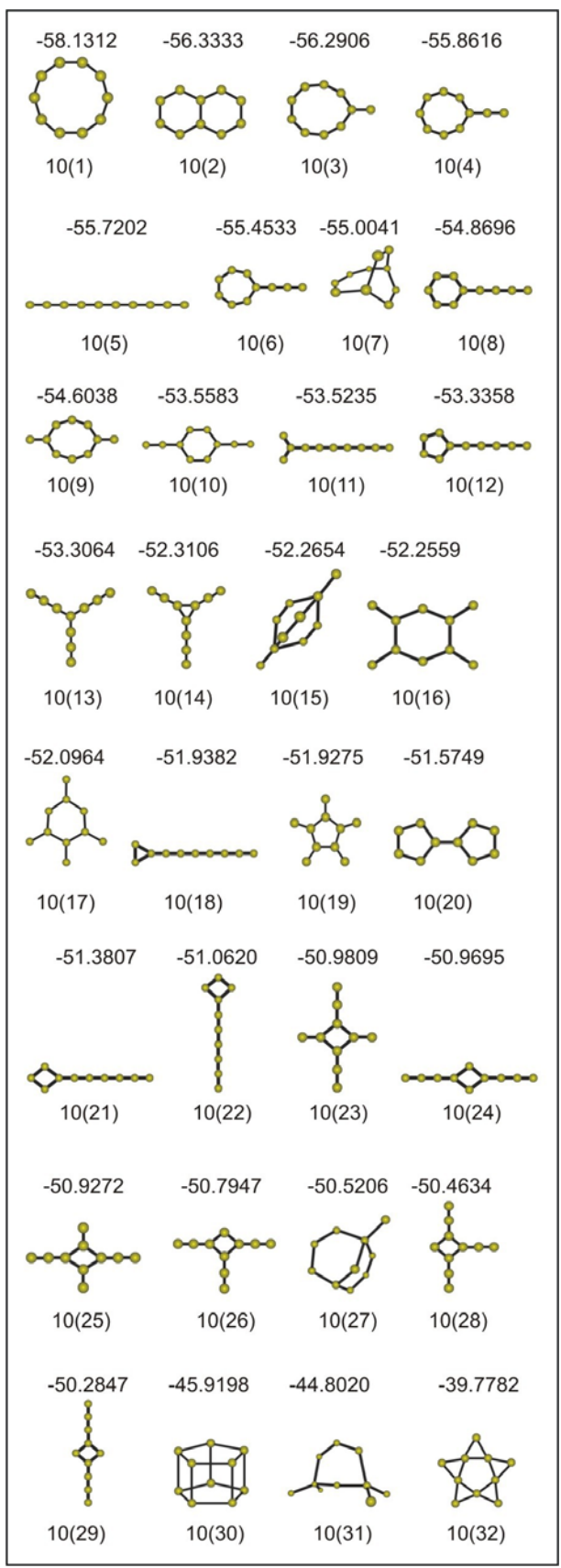

**FIG. 6**

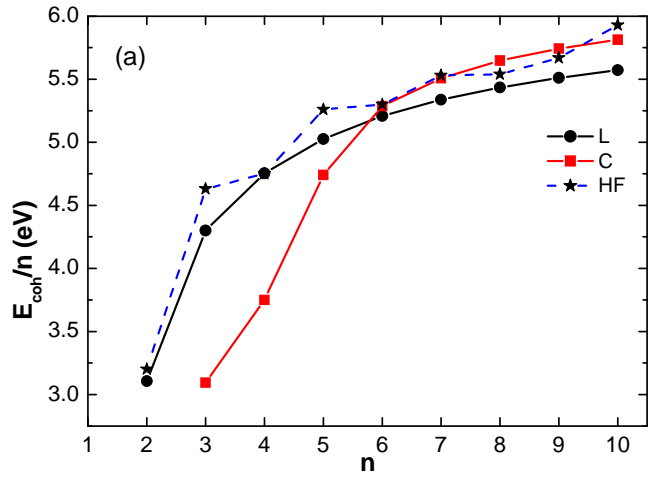

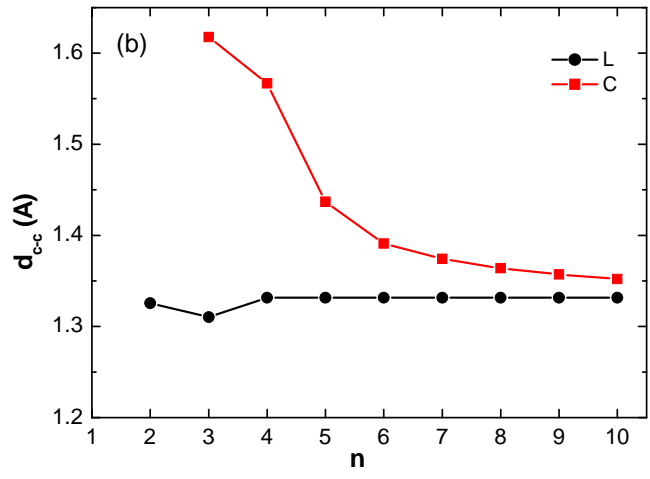

**FIG. 7**

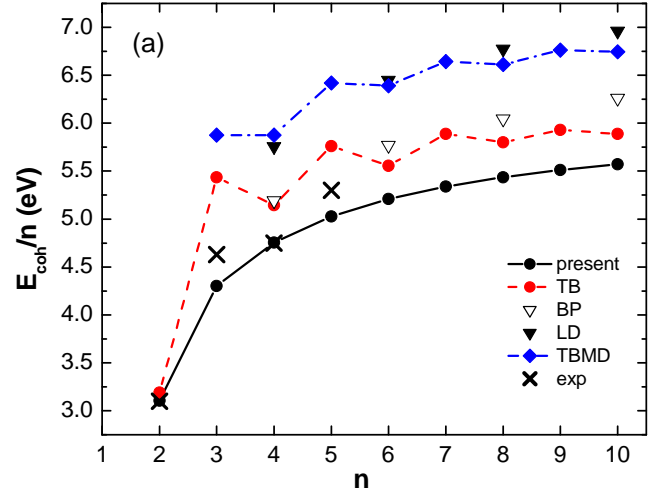

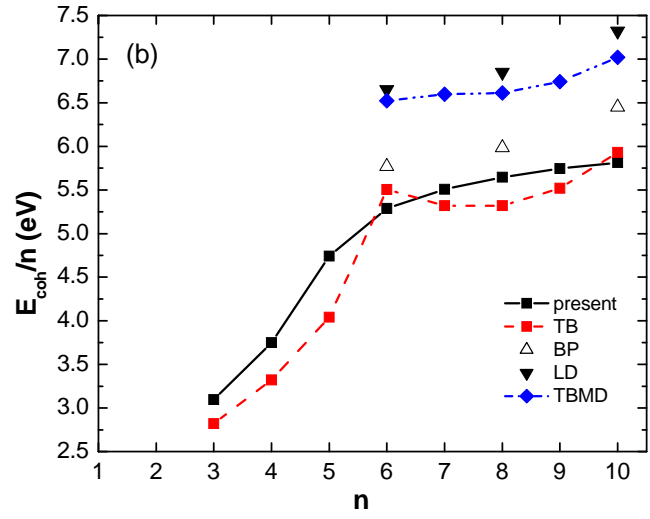

**FIG. 8**